\documentclass[twocolumn,aps,superscriptaddress,prb,longbibliography]{revtex4-1}
\usepackage{graphicx}
\usepackage{amssymb}
\usepackage{amsmath}
\usepackage{bm}
\usepackage{color}

\def\Journal #1,#2,#3,#4#5#6#7{#1 {\bf #2}, #3 (#4#5#6#7)}

\def\Vec{\mathbf}
\def\GVec#1{\mbox{\boldmath $#1$}}

\def\lsim{\, \lower -0.3ex \hbox{$<$} \kern -0.75em \lower 0.7ex \hbox{$\sim$} \,}
\def\gsim{\, \lower -0.3ex \hbox{$>$} \kern -0.75em \lower 0.7ex \hbox{$\sim$} \,}

\begin{document}

\title{Continuum model for relaxed twisted bilayer graphenes and moir\'{e} electron-phonon interaction
}
\author{Mikito Koshino}
\thanks{koshino@phys.sci.osaka-u.ac.jp}
\affiliation{Department of Physics, Osaka University,  Toyonaka 560-0043, Japan}

\author{Nguyen N. T. Nam}
\affiliation{Mathematics for Advanced Materials Open Innovation Lab (MathAM-OIL), AIST, Sendai 980-8577, Japan}

\begin{abstract}
We construct an analytic continuum model to describe the electronic structure 
and the electron-phonon interaction in twisted bilayer graphenes with arbitrary lattice deformation.
Starting from the tight-binding model, we derive the interlayer Hamiltonian in the presence of  general lattice displacement,
and obtain a long-wavelength continuum expression for smooth deformation.
We show that the continuum model correctly describes the 
band structures of the lattice-relaxed twisted bilayer graphenes. 
We apply the formula to the phonon vibration,
and derive an explicit expression of the electron-phonon matrix elements between the moir\'{e} band states and 
the moir\'{e} phonon modes. 
By numerical calculation, we find that the electron-phonon coupling and phonon mediated electron-electron interaction
are significantly enhanced in low twist angles due to the superlattice hybridization.
\end{abstract}

\maketitle

\section{Introduction}
\label{sec_intro}

Twisted bilayer graphene (TBG) is a rotationaly-stacked graphene bilayer system
governed by a nanoscale moir\'{e} interference pattern between the mismatched layers.
\cite{lopes2007graphene,mele2010commensuration,trambly2010localization,shallcross2010electronic,
morell2010flat,bistritzer2011moirepnas,kindermann2011local,xian2011effects,PhysRevB.86.155449,moon2012energy,de2012numerical}
The physical properties of TBG are sensitive to lattice distortion,
because a slight change in the atomic lattice is magnified to a big deformation in the moir\'{e} pattern,
resulting in a significant influence on the electronic system.
The actual atomic configuration of the real TBG is not a simple stack of rigid graphene layers, but
it contains a triangular  AB/BA-stacking domain structure
as a consequence of spontaneous lattice relaxation.  \cite{popov2011commensurate, brown2012twinning,lin2013ac,alden2013strain,uchida2014atomic,
van2015relaxation,dai2016twisted, jain2016structure,nam2017lattice,carr2018relaxation, lin2018shear,yoo2019atomic,guinea2019continuum}.
Such a structural deformation strongly affects
the electronic band structure. \cite{nam2017lattice, lin2018shear, koshino2018maximally, yoo2019atomic,guinea2019continuum,crucial2019lucignano},

The moir\'{e} pattern plays an important role also in the lattice vibration.
While the wide-range phonon spectrum of TBG resembles that of regular AB-stacked bilayer graphene
\cite{jiang2012acoustic,cocemasov2013phonons,ray2016electron,choi2018strong}, 
the detailed phonon structure is 
actually subject to a significant influence from the moir\'{e} effect. \cite{angeli2019valley, koshino2019moire}
In the low-energy acoustic branch, in particular, the phonon spectrum
is reconstructed into superlattice minibands ruled by the moir\'{e} period,
where the eigen phonon modes can be regarded as
effective vibration modes of the triangular domain structure. \cite{koshino2019moire}
These moir\'{e} phonon modes are expected to strongly interact with the flat band
electronic states and affect the correlated phenomena. \cite{cao2018unconventional,cao2018mott,yankowitz2019tuning}
The electron-phonon interaction in TBG is theoretically studied
by considering bare phonons without superlattice modulation \cite{ray2016electron, choi2018strong,wu2018theory, wu2019phononinduced,lian2019twisted},
while the effect of the reconstructed moir\'{e} phonon modes is not well understood.

The purpose of this work is to develop a general continuum model for TBG
with arbitrary lattice deformation, including lattice relaxation and phonon vibrations.
The electronic properties of TBGs have been mostly studied
by the continuum model, \cite{lopes2007graphene,bistritzer2011moirepnas,kindermann2011local,PhysRevB.86.155449,moon2013opticalabsorption,koshino2015interlayer,koshino2015electronic,weckbecker2016lowenergy}
which can efficiently describe the band structure by capturing the long-wave components.
However, the continuum model was originally derived for rigid TBGs,
and the extension to deformed TBGs is necessary to study the above issues.
Inclusion of the relaxation effect in the continuum model has been considered in very recent works.
\cite{koshino2018maximally,carr2019exact,guinea2019continuum,fang2019angle,fleischmann2019perfect,balents2019general,ochoa2019moire}

In this paper, we construct an analytic continuum Hamiltonian for TBGs under arbitrary lattice deformation.
Starting from the tight-binding model, 
we derive the interlayer Hamiltonian matrix
as an analytic function of lattice displacement vectors.
We then apply the continuum model to the relaxed TBGs with domain structure\cite{nam2017lattice}, 
and demonstrate the energy band of the original tight-binding model
is correctly reproduced. 
Lastly we apply the formula to the phonon vibration,
and obtain the electron-phonon matrix elements for the moir\'{e} phonon modes in relaxed TBGs.

The paper is organized as follows.
In Sec.\ \ref{sec_continuum}, we present a general theoretical treatment 
to describe the electronic coupling between deformed graphene layers,
and write down the interlayer matrix elements
for arbitrary lattice displacement [Eq.\ (\ref{eq_umat_3d})].
We then obtain its long-wavelength expression in a simple form [Eq.\ (\ref{eq_umat_3d_approx2})],
which is valid for small twist angles and smooth deformation.
In Sec.\ \ref{sec_band}, we apply the continuum Hamiltonian to relaxed TBGs
and calculate the band structure.
In Sec.\ \ref{sec_e-p}, we derive the quantized Hamiltonian for the lattice vibration of the relaxed TBG,
and then obtain the explicit formula for the matrix elements between the electrons
and the quantized moir\'{e} phonons.


\section{Continuum Hamiltonian of TBG with lattice deformation}
\label{sec_continuum}

\subsection{Lattice geometry}

Let us consider a TBG with twist angle $\theta$,
and define its non-distorted structure by bilayer of intrinsic monolayer graphenes 
stacked with the graphite's interlayer spacing $d_0 = 0.335$nm
and in-plane rotation by $\mp \theta/2$ for layer 1 and 2, respectively.
We take $x,y$ axes on the graphene layer, and $z$ axis perpendicular to it.
The primitive lattice vectors of layer 1 is defined by $\Vec{a}^{(1)} _i =R(- \theta/2)\Vec{a}_i$
and those of layer 2 by  $\Vec{a}^{(2)} _i =R(\theta/2) \Vec{a}_i$ ($i=1,2$),
where  $\Vec{a}_1 = a(1,0)$ and $\Vec{a}_2= a(1/2,\sqrt{3}/2)$ are the lattice vectors before the rotation, 
$a \approx 0.246\,\mathrm{nm}$ is the graphene's lattice constant,
and $R$ is the rotation matrix.
The unit cell area of monolayer graphene is given by $S_0 = |\Vec{a}_1\times \Vec{a}_2|$.
The primitive reciprocal lattice vectors of layer $l$ are $\Vec{a}_i^{*(l)} =R(\mp \theta/2)\Vec{a}^*_i$,
where $\Vec{a}^*_1 = (2\pi/a)(1,-1/\sqrt{3})$ and $\Vec{a}^*_2=(2\pi/a)(0,2/\sqrt{3})$
are those for non-rotated graphene.

Each graphene layer contains two sublattices
labeled by $X = A, B$ in its own unit cell.
In the absence of the lattice distortion,
the positions of sublattice $X$ on layer $l$ are given by \begin{align}
&\Vec{R}^{(l)}_{X}=m_1\Vec{a}^{(l)}_{1}+m_2\Vec{a}^{(l)}_{2}+\GVec{\tau}^{(l)}_{X}
\end{align}
Here $m_1$ and $m_2$ are integers, 
and  $\GVec{\tau}^{(l)}_{X}$ is the relative sublattice position
inside the unit cell, which are given by
$\GVec{\tau}^{(1)}_{A}= 0$,  $\GVec{\tau}^{(1)}_{B}= (\Vec{a}^{(1)}_1-2\Vec{a}^{(1)}_2)/3$, 
$\GVec{\tau}^{(2)}_{A}= d_0\,\Vec{e}_z$ and 
$\GVec{\tau}^{(2)}_{B}=  (\Vec{a}^{(2)}_1-2\Vec{a}^{(2)}_2)/3 + d_0\,\Vec{e}_z$.
Here $\Vec{e}_z$ is the unit vector in $z$ direction,
and $d_0$ is the interlayer spacing in the absence of distortion,
which is set to the interlayer distance of graphite, 0.334nm.

\begin{figure}
\begin{center}
\leavevmode\includegraphics[width=	0.9\hsize]{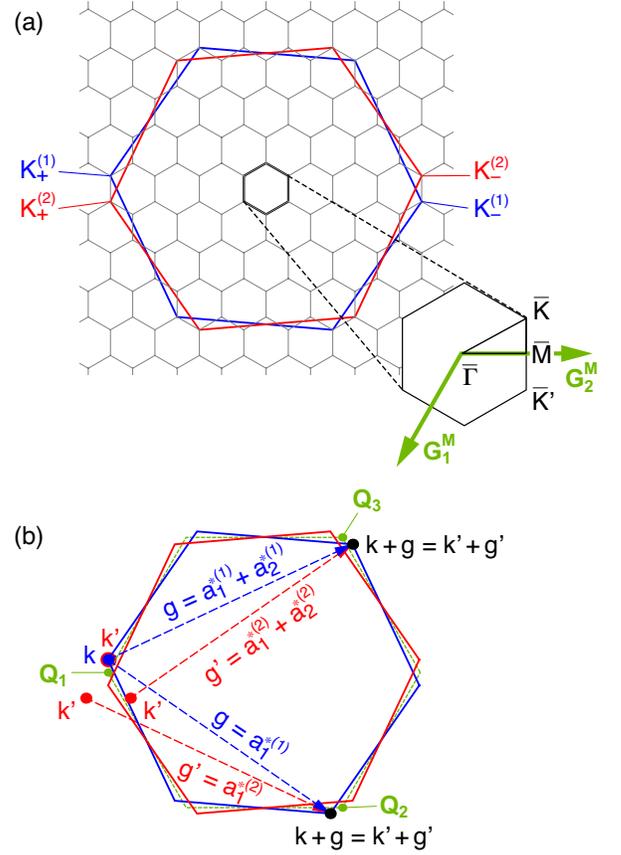}
\end{center}
\caption{
(a) Brillouin zone folding in TBG with $\theta = 9.43^\circ$.
Two large hexagons represent the first Brillouin zones of graphene layer 1 and 2, 
and the small hexagon is the moir\'{e} Brillouin zone of TBG.
(b) Schematic diagram to illustrate the interlayer coupling
from an initial $\Vec{k}$ of layer 1(taken as $\Vec{K}^{(1)}_+$) 
to three $\Vec{k}'$ points of layer 2 in a non-distorted TBG (see the text).
}
\label{fig_BZ}
\end{figure}

In a small angle TBG, a slight mismatch of the lattice periods gives rise to a 
moir\'{e} interference pattern. The reciprocal lattice vectors for the moir\'{e} pattern is given by
$ \Vec{G}^{\rm M}_{i}  =  \textbf{a}^{*(1)}_i - \textbf{a}^{*(2)}_i \, (i=1,2).$
The real-space lattice vectors $\Vec{L}^{\rm M}_{j}$ can be obtained from
$\Vec{G}^{\rm M}_i\cdot\Vec{L}^{\rm M}_{j} = 2\pi\delta_{ij}$.
The moir\'{e} lattice constant $L_{\rm M} = | \Vec{L}^{\rm M}_{i}|$ is given by
$L_{\rm M} =  a/[2\sin (\theta/2)]$.
Figure \ref{fig_BZ}(a) shows the folding of the Brillouin zone, 
where two large hexagons represent the first Brillouin zones of layer 1 and 2, 
and the small hexagon is the moir\'{e} Brillouin zone of TBG.
The graphene's Dirac points (the band touching points) 
are located at  $\Vec{K}^{(l)}_\xi = R(\mp \theta/2) \Vec{K}_\xi$
with $\mp$ for layer 1 and 2, respectively,
where $\Vec{K}_\xi = -\xi [2\Vec{a}^*_1+\Vec{a}^*_2]/3 = -\xi(2\pi/a)(2/3,0)$
is the Dirac points before rotation and $\xi=\pm 1$ is the valley index.
In a small twist angle,
$\Vec{K}^{(1)}_\xi$ and $\Vec{K}^{(2)}_\xi$ of the same valley $\xi$ are displaced 
only by a short distance of the order of $\Vec{G}_i^M$.

\subsection{Interlayer Hamiltonian under general lattice distortion}

Here we derive the general formula, Eq.\ (\ref{eq_umat_3d}),
which describes the interlayer coupling between the Bloch states in TBG in
the presence of arbitrary lattice distortion. 
We define $\Vec{u}_X^{(l)}(\Vec{R})$
as displacement vector of an atomic site of sublattice $X$ on layer $l$, which is originally located 
$\Vec{R} (\in \Vec{R}^{(l)}_X)$.
The $\Vec{u}_X^{(l)}$ can be a three-dimensional vector in general.
We expand the displacement vector in the Fourier series as
\begin{align}
& \Vec{u}_X^{(l)}(\Vec{R}) = \sum_{\Vec{q}} \Vec{u}^{(l)}_{X,\Vec{q}} e^{i \Vec{q}\cdot\Vec{R}},
\label{eq_uq_FT}
\end{align}
where the summation in $\Vec{q}$ is taken over two-dimensional wave numbers.

Let $|\Vec{R}\rangle$ be a carbon $p_z$ orbital at site $\Vec{R}$.
We define the Bloch bases under the lattice distortion as
\begin{align}
	& |\Vec{k},X, l\rangle \equiv 
	\frac{1}{\sqrt{N}}\sum_{\Vec{R} \in \Vec{R}^{(l)}_{X}} e^{i\Vec{k}\cdot\Vec{R}}
	|\Vec{R} + \Vec{u}_X^{(l)}(\Vec{R}) \rangle
	\label{eq_bloch_base}
\end{align}
where $\Vec{k}$ is a two-dimensional Bloch wave vector, 
and $N=S/S_0$ is the number of graphene's unit cells per layer
in the total system area $S$.
We assume the interlayer hopping from $|\Vec{R}^{(1)}\rangle$ to $|\Vec{R}^{(2)}\rangle$ is given by
 \begin{align}
\langle \Vec{R}^{(2)} |U| \Vec{R}^{(1)} \rangle = -T(\Vec{R}^{(2)} -\Vec{R}^{(1)}),
 \end{align}
and define the three-dimensional Fourier transform as
\begin{align}
t(\Vec{p}) = \frac{1}{S_0 d_0}  
\int d^3 r\,\,\, T(\Vec{r}) e^{-i \Vec{p}\cdot\Vec{r}},
\end{align}
where $\Vec{p}=(p_x,p_y,p_z)$ is a three dimensional wave vector.

In the actual band calculation in Sec.\ \ref{sec_band}, 
we will use the standard Slater-Koster form for $T(\Vec{R})$,
\begin{eqnarray}
&& -T(\Vec{R}) = 
V_{pp\pi}\left[1-\left(\frac{\Vec{R}\cdot\Vec{e}_z}{R}\right)^2\right]
+ V_{pp\sigma}\left(\frac{\Vec{R}\cdot\Vec{e}_z}{R}\right)^2,
\nonumber \\
&& V_{pp\pi} =  V_{pp\pi}^0 e^{- (R-a_0)/r_0},
\,\, V_{pp\sigma} =  V_{pp\sigma}^0  e^{- (R-d_0)/r_0}.
\label{eq_transfer_integral}
\end{eqnarray}
Here $\Vec{e}_z$ is the unit vector perpendicular to the graphene plane,
$a_0 = a/\sqrt{3} \approx 0.142\,\mathrm{nm}$ is the distance of
neighboring $A$ and $B$ sites,
and the parameter $V_{pp\pi}^0 \approx -2.7\,\mathrm{eV}$ is the transfer integral between 
the nearest-neighbor atoms on graphene, and $V_{pp\sigma}^0 \approx 0.48\,\mathrm{eV}$ is the transfer integral
between vertically located atoms on the neighboring layers of graphite,
$r_0 = 0.184 a$ is the decay length of the transfer integral. \cite{moon2013opticalabsorption}

The interlayer matrix element between the Bloch bases is written as
\begin{align}
& \langle \Vec{k}',X',2 |U| \Vec{k},X,1 \rangle 
\nonumber\\
&=
 - \frac{1}{N}
\sum_{\Vec{R} \in \Vec{R}^{(1)}_{X}} 
\sum_{\Vec{R}' \in \Vec{R}^{(2)}_{X'}} \, e^{i\Vec{k}\cdot\Vec{R}-i\Vec{k}'\cdot\Vec{R}'}
\nonumber\\
&\qquad\quad
\times T[\Vec{R}' + \Vec{u}_{X'}^{(2)}(\Vec{R}') -\Vec{R} - \Vec{u}_X^{(1)}(\Vec{R})]
\nonumber\\
&
= - \frac{1}{N} 
\frac{S_0 d_0}{(2\pi)^3} \int  d^3 p
\,\, t(\Vec{p})
\sum_{\Vec{R} \in \Vec{R}^{(1)}_{X}} 
e^{i(\Vec{k}-\Vec{p})\cdot \Vec{R}  - i \Vec{p}\cdot \Vec{u}_X^{(1)}(\Vec{R})}
\nonumber\\
&\qquad\qquad\qquad\times
\sum_{\Vec{R}' \in \Vec{R}^{(2)}_{X'}}
 e^{-i(\Vec{k}'-\Vec{p})\cdot \Vec{R}'  + i \Vec{p}\cdot \Vec{u}_{X'}^{(2)}(\Vec{R}')}
.\label{eq_umat_1}
 \end{align}
We replace $\Vec{u}_X^{(1)}(\Vec{R})$ and  $\Vec{u}_{X'}^{(2)}(\Vec{R}')$ 
with its Fourier transform in Eq.\ (\ref{eq_uq_FT}),
and expand the exponential functions such
 as $\exp(i \Vec{p}\cdot \Vec{u}_{\Vec{q}} e^{i\Vec{q}\cdot\Vec{R}})$
in a Taylor series as
\begin{align}
\exp(i \Vec{p}\cdot \Vec{u}_{\Vec{q}} \, e^{i\Vec{q}\cdot\Vec{R}})
= \sum_{n=0}^\infty \frac{1}{n!}(i \Vec{p}\cdot \Vec{u}_{\Vec{q}})^n e^{i n \Vec{q}\cdot\Vec{R}}.
\label{eq_taylor}
\end{align}
Then we can take the summation over the lattice points by using
\begin{align}
\sum_{\Vec{R} \in \Vec{R}^{(l)}_{X}} 
e^{i\Vec{p} \cdot\Vec{R}}
= N 
 \sum_{\Vec{g}}
 e^{i(\Vec{g} + p_z \Vec{e}_z)\cdot\mbox{\boldmath \scriptsize $\tau$}^{(l)}_{X}}
  \delta_{\Vec{p}_\parallel,\Vec{g}},
\end{align}
where $\Vec{p}_\parallel$ is $xy$-component of $\Vec{p}$, 
and the summation in $\Vec{g}$ is taken over all the reciprocal lattice vectors 
$\Vec{g}=m_1\Vec{a}^{*(l)}_1+m_2\Vec{a}^{*(l)}_2$.

Using these, we obtain a formula 
\begin{align}
& \langle \Vec{k}',X',2 |U| \Vec{k},X,1 \rangle
\nonumber\\
&
= \sum_{\Vec{g},\Vec{g}'}
\sum_{n_1,n_2,\cdots}\sum_{n'_1,n'_2,\cdots}
 \Gamma_{(n_1,n_2,\cdots)}^{(n'_1,n'_2,\cdots)}(\Vec{Q})
\,\, e^{-i\Vec{g}\cdot\mbox{\boldmath \scriptsize $\tau$}^{(1)}_{X}
	+i\Vec{g}'\cdot\mbox{\boldmath \scriptsize $\tau$}^{(2)}_{X'}}
 \nonumber\\
&
\qquad\times \delta_{
 \Vec{k}+\Vec{g}+n_1\Vec{q}_1+n_2\Vec{q}_2+\cdots, \,
 \Vec{k}'+\Vec{g}'-n'_1\Vec{q}_1-n'_2\Vec{q}_2+\cdots
 },
 \label{eq_umat_3d}
\end{align}
where $\Vec{g}=m_1\Vec{a}^{*(1)}_1+m_2\Vec{a}^{*(1)}_2$ and
$\Vec{g}'=m'_1\Vec{a}^{*(2)}_1+m'_2\Vec{a}^{*(2)}_2$,
$n_i, n'_i = 0,1,2,\cdots$, 
$\Vec{Q} = \Vec{k}+\Vec{g}+n_1\Vec{q}_1+n_2\Vec{q}_2+\cdots$,
and
\begin{align}
&\Gamma_{(n_1,n_2,\cdots)}^{(n'_1,n'_2,\cdots)}
(\Vec{Q})
= -\frac{d_0}{2\pi}  \int_{-\infty}^\infty dp_z 
t(\Vec{Q}') e^{i p_z d_0}
\nonumber\\
&\quad\times
\frac{[-i\Vec{Q}'\cdot\Vec{u}^{(1)}_{X,\Vec{q}_1}]^{n_1}}{n_1!}
\frac{[-i\Vec{Q}'\cdot\Vec{u}^{(1)}_{X,\Vec{q}_2}]^{n_2}}{n_2!} \cdots
\nonumber\\
&\quad\times
\frac{[+i\Vec{Q}'\cdot\Vec{u}^{(2)}_{X',\Vec{q}_1}]^{n'_1}}{n'_1!}
\frac{[+i\Vec{Q}'\cdot\Vec{u}^{(2)}_{X',\Vec{q}_2}]^{n'_2}}{n'_2!} \cdots,
\label{eq_gamma_func}
\end{align}
with $\Vec{Q}' = \Vec{Q}+p_z\Vec{e}_z$.
Here $\{\Vec{q}_1,\Vec{q}_2,\cdots\}$ is a set of the two-dimensional wave numbers 
in which $\Vec{u}^{(l)}_{X,\Vec{q}}$ has finite Fourier amplitudes.
Eq.\ (\ref{eq_umat_3d}) simply means that
the Bloch states of the layer 1 and layer 2 are coupled when 
\begin{align}
&\Vec{k}+\Vec{g}+n_1\Vec{q}_1+n_2\Vec{q}_2+\cdots
\nonumber\\
&\quad =
 \Vec{k}'+\Vec{g}'-n'_1\Vec{q}_1-n'_2\Vec{q}_2+\cdots (=\Vec{Q}),
\label{eq_umklapp}
\end{align}
and its coupling amplitude is given by 
$\Gamma_{(n_1,n_2,\cdots)}^{(n'_1,n'_2,\cdots)}(\Vec{Q})$.
The higher order terms in $n_i$ and $n'_i$ in Eq.\ (\ref{eq_gamma_func}) quickly decay as long as
$|\Vec{Q}'\cdot\Vec{u}^{(l)}_{X,\Vec{q}_j}|$ is sufficiently small,
and it is the case in TBGs considered below.

When the displacement vector is along in-plane direction (i.e., $u_z^{(l)} = 0$), in particular,
Eq.\ (\ref{eq_gamma_func}) is reduced to
\begin{align}
&\Gamma_{(n_1,n_2,\cdots)}^{(n'_1,n'_2,\cdots)}(\Vec{Q})
= -t_\parallel(\Vec{Q};\, d_0) 
\nonumber\\
&\quad\times
\frac{[-i\Vec{Q}\cdot\Vec{u}^{(1)}_{X,\Vec{q}_1}]^{n_1}}{n_1!}
\frac{[-i\Vec{Q}\cdot\Vec{u}^{(1)}_{X,\Vec{q}_2}]^{n_2}}{n_2!} \cdots
\nonumber\\
&\quad\times
\frac{[+i\Vec{Q}\cdot\Vec{u}^{(2)}_{X',\Vec{q}_1}]^{n'_1}}{n'_1!}
\frac{[+i\Vec{Q}\cdot\Vec{u}^{(2)}_{X',\Vec{q}_2}]^{n'_2}}{n'_2!} \cdots,
\label{eq_gamma_func_in-plane}
\end{align}
where
\begin{align}
t_\parallel(\Vec{Q};\, z) 
&=  
\frac{d_0}{2\pi}  \int_{-\infty}^\infty dp_z\,\, t(\Vec{Q}+p_z\Vec{e}_z) e^{i p_z z}
\nonumber
\\
& = \frac{1}{S_0}\int d^2 r \,\, T(\Vec{r} + z \Vec{e}_z) 
e^{-i \Vec{Q} \cdot\Vec{r}},
\label{eq_t_para}
\end{align}
is the two-dimensional Fourier transform of $T(\Vec{r})$ on a plane parallel to $xy$ at fixed height $z$.

In the absence of the displacement (i.e., $\Vec{u}^{(l)}_X = 0$),
the terms in the Taylor series remain only when $n_j = n'_j = 0$,
and Eq.\ (\ref{eq_umat_3d}) becomes
\begin{align}
& \langle \Vec{k}',X',2 |U| \Vec{k},X,1 \rangle
\nonumber\\
&= -\sum_{\Vec{g},\Vec{g}'} 
t_\parallel(\Vec{k}+\Vec{g};\, d_0)
\, e^{-i\Vec{g}\cdot\mbox{\boldmath \scriptsize $\tau$}^{(1)}_{X}
	+i\Vec{g}'\cdot\mbox{\boldmath \scriptsize $\tau$}^{(2)}_{X'}}
 \delta_{ \Vec{k}+\Vec{g},\Vec{k}'+\Vec{g}'},
 \label{eq_umat_3d_norelax}
\end{align}
which is the interlayer Hamiltonian of non-distorted TBG in the earlier works.
\cite{bistritzer2011moirepnas, koshino2015interlayer}

\subsection{Continuum Hamiltonian for small twist angles}

Eq.\ (\ref{eq_umat_3d}) is the general formula which works for any twist angles
with arbitrary displacement vectors.
Here we will derive a long-range approximate form, Eq.\ (\ref{eq_umat_3d_approx2}),
which is valid for small twist angles and long-range displacement.
In the following, we assume that the moir\'{e} period much greater than the atomic scale,
and also that $\textbf{u}^{(l)}_A(\Vec{r}) = \textbf{u}^{(l)}_B(\Vec{r}) = \textbf{u}^{(l)}(\Vec{r})$,
where $ \textbf{u}^{(l)}$ is a smoothly varying function compared to the atomic scale.

We first consider the non-distorted case, $\Vec{u}^{(l)}=0$.
In Eq.\ (\ref{eq_umat_3d_norelax}),
the Bloch states at $\Vec{k}$ (layer 1) and $\Vec{k}'$ (layer 2) are mixed when 
$\Vec{k}+\Vec{g} = \Vec{k}'+\Vec{g}'$, and then the coupling amplitude is 
given by $t_\parallel(\Vec{k}+\Vec{g};\, d_0)$.
Here only a few terms are relevant in the summation over $\Vec{g}$ and $\Vec{g}'$,
because the function $t_\parallel(\Vec{k}+\Vec{g};\, d_0)$ quickly decays
for large $|\Vec{k}+\Vec{g}|$.
When we start from $\Vec{k} \approx \Vec{K}_\xi$ to consider a low-energy state near the Fermi energy,
the dominant coupling occurs in three cases
$(\Vec{g},\Vec{g}') = (\Vec{0},\Vec{0}), \xi (\Vec{a}^{*(1)}_1,\Vec{a}^{*(2)}_1),
 \xi (\Vec{a}^{*(1)}_1+\Vec{a}^{*(1)}_2,\Vec{a}^{*(2)}_1+\Vec{a}^{*(2)}_2)$.
Figure \ref{fig_BZ}(b) shows the positions of $\Vec{k}$, $\Vec{k}'$ and $\Vec{k}+\Vec{g}$
for an initial vector $\Vec{k} = \Vec{K}^{(1)}_+$.
The corresponding $\Vec{k}+\Vec{g}$ is close to 
three equivalent corner points of  the first Brillouin zone of non-rotated graphene,
 \begin{align}
&\Vec{Q}_1 = \Vec{K}_\xi, \,\, \Vec{Q}_2 = \Vec{K}_\xi +\xi \Vec{a}^*_1,  
\,\, \Vec{Q}_3 = \Vec{K}_\xi + \xi(\Vec{a}^*_1 + \Vec{a}^*_2).
\end{align}
By neglecting a small shift, we can replace $\Vec{k}+\Vec{g}$ with $\Vec{Q}_j$
in $t_\parallel(\Vec{k}+\Vec{g};\, d_0)$  of Eq.\ (\ref{eq_umat_3d_norelax}).
This gives the widely-used continuum model for the non-distorted TBG. \cite{bistritzer2011moirepnas}

The same approximation can be used
in the presence of a long-range lattice distortion $\Vec{u}^{(l)}$.
When $\Vec{q}$'s in $\Vec{u}^{(l)}_{\Vec{q}}$ 
are much shorter than $1/a$ (i.e.,   $\Vec{u}^{(l)}(\Vec{r})$ is smoothly varying compared to $a$),
we can neglect a small shift $n_1\Vec{q}_1+n_2\Vec{q}_2+\cdots$ in $\Vec{Q}$
in the coupling amplitude $\Gamma_{(n_1,n_2,\cdots)}^{(n'_1,n'_2,\cdots)}(\Vec{Q})$.
Then we can replace $\Vec{Q}$ with the above three $\Vec{Q}_j$'s in Eq.\ (\ref{eq_umat_3d}) to obtain,
\begin{align}
& \langle \Vec{k}',X',2 |U| \Vec{k},X,1 \rangle
\nonumber\\
&\qquad
= \sum_{j=1}^3
\sum_{n_1,n_2,\cdots}\sum_{n'_1,n'_2,\cdots}
 \Gamma_{(n_1,n_2,\cdots)}^{(n'_1,n'_2,\cdots)}(\Vec{Q}_j)
M^j_{X'X}
 \nonumber\\
&\qquad\qquad
\times \delta_{
 \Vec{k}',  \Vec{k}+\delta\Vec{k}_j+(n_1+n'_1)\Vec{q}_1+(n_2+n'_2)\Vec{q}_2+\cdots
 },
\label{eq_umat_3d_approx1}
\end{align}
where $\delta \Vec{k}_j (= \Vec{g}-\Vec{g}')$ is given by
\begin{align}
&\delta \Vec{k}_1 = \Vec{0}, \,\, \delta\Vec{k}_2 = \xi \Vec{G}^{M}_1,  
\,\,  \delta\Vec{k}_2 = \xi (\Vec{G}^{M}_1 + \Vec{G}^{M}_2),
\end{align}
and $M^j_{X'X} (= e^{-i\Vec{g}\cdot\mbox{\boldmath \scriptsize $\tau$}^{(1)}_{X}
	+i\Vec{g}'\cdot\mbox{\boldmath \scriptsize $\tau$}^{(2)}_{X'}}$) is 
\begin{align}
M^1 = \begin{pmatrix}
1 & 1 \\
1 & 1
\end{pmatrix},
M^2 = 
\begin{pmatrix}
1 & \omega^{-\xi} \\
\omega^\xi & 1
\end{pmatrix},
\,\,
M^3 = 
\begin{pmatrix}
1 & \omega^\xi \\
\omega^{-\xi} & 1
\end{pmatrix},
\,\,
\end{align}
where $\omega = e^{2\pi i /3}$ and $M^j$ stands for 
\begin{align}
& M^j = 
\begin{pmatrix}
M^j_{AA} & M^j_{AB} \\
M^j_{BA} & M^j_{BB}
\end{pmatrix}.
\end{align}

In the real space representation, Eq.\ (\ref{eq_umat_3d_approx1}) is simply expressed as
\begin{align}
& \langle \Vec{k}',X',2 |U| \Vec{k},X,1 \rangle
= \frac{1}{S} \int d^2 r
e^{i(\Vec{k}-\Vec{k}')\cdot\Vec{r}}
U_{X'X} (\Vec{r}),
\end{align}
where
\begin{align}
U_{X'X} (\Vec{r}) = 
&-\sum_{j=1}^3
M^j_{X'X} \,\,
t_\parallel [\Vec{Q}_j; d_0 + u^-_{z}(\Vec{r})]
\nonumber\\
&\qquad
\times \exp[i \Vec{Q}_j\cdot \Vec{u}^-(\Vec{r}) + i \delta\Vec{k}_j\cdot\Vec{r}].
 \label{eq_umat_3d_approx2}
\end{align}
where $\Vec{u}^{-} = \Vec{u}^{(2)}  - \Vec{u}^{(1)}$
 is the interlayer asymmetric displacement vector between the two layers.
From Eq.\ (\ref{eq_umat_3d_approx1}) to  Eq.\ (\ref{eq_umat_3d_approx2}),
we used 
$\delta_{\Vec{k}',  \Vec{k}+\Vec{q}}  = (1/S)\int d^2r\, e^{i(\Vec{k}+ \Vec{q} -\Vec{k}' )}$,
and applied Eq.\ (\ref{eq_taylor}) inversely, and finally used Eq.\ (\ref{eq_t_para}) for the integral in $p_z$.
Note that the interlayer Hamiltonian matrix of $U$ only depends on the asymmetric displacement $\Vec{u}^{-}$,
but not on the symmetric part $\Vec{u}^{+} = \Vec{u}^{(2)}  + \Vec{u}^{(1)}$.

In in-plane distortion (i.e., $u_z^{(l)} = 0$),  particularly,
Eq.\ (\ref{eq_umat_3d_approx2}) becomes
\begin{align}
U_{X'X} (\Vec{r}) = 
t_0 \sum_{j=1}^3
M^j_{X'X} \exp[i \Vec{Q}_j\cdot \Vec{u}^-(\Vec{r}) + i \delta\Vec{k}_j\cdot\Vec{r}],
\label{eq_umat_in-plane}
\end{align}
where  
$t_0 = -t_\parallel [\Vec{Q}_j; d_0].$
Note that $t_0$ is independent of $j$
because $|\Vec{Q}_1| = |\Vec{Q}_2| = |\Vec{Q}_3| = 4\pi/(3a)$, and
$t_\parallel (\Vec{q}; d_0)$ is circularly symmetric.
In the present hopping model of Eq.\  (\ref{eq_transfer_integral}), we have $t_0 = 0.104$ eV.

The total continuum Hamiltonian of TBG for the valley $\xi$ can be written in a 4 $\times $ 4 matrix for the basis of $(A_1,B_1,A_2,B_2)$ as
 \begin{eqnarray}
	{H} = 
	\begin{pmatrix}
		H_1 & U^\dagger \\
		U & H_2
	\end{pmatrix}.
	\label{eq_H_eff}
\end{eqnarray}
Here $U$ is given by Eq.\ (\ref{eq_umat_3d_approx2}).
The $H_l\, (l=1,2)$ is the intralayer Hamiltonian of layer $l$,
which is given by the two-dimensional Weyl equation centered at $\Vec{K}^{(l)}_\xi$ point, 
\begin{align}
& H_l(\Vec{k}) = - \hbar v \left[
R(\pm \theta/2)
\left({\Vec{k}}+\frac{e}{\hbar}\Vec{A}^{(l)}-\Vec{K}^{(l)}_\xi
\right)
\right]
 \cdot (\xi \sigma_x, \sigma_y),
	\label{eq_intra}
\end{align}
where $\pm$ is for $l=1$ and 2, respectively, $v$ is the graphene's band velocity,
and $\sigma_x, \sigma_y$ are the Pauli matrices in the sublattice space $(A,B)$.
We take $\hbar v /a = 2.1354$ eV.\cite{moon2013opticalabsorption}
The $\Vec{A}^{(l)}$ is the pseudo-vector potential induced by the lattice strain, 
which given by \cite{suzuura2002phonons,pereira2009strain,guinea2010energy}
\begin{eqnarray}
A^{(l)}_x &=& \xi\frac{3}{4}\frac{\beta\gamma_0}{ev}[u^{(l)}_{xx}-u^{(l)}_{yy}], 
\nonumber\\
A^{(l)}_y &=&  \xi\frac{3}{4}\frac{\beta\gamma_0}{ev} [-2u^{(l)}_{xy}].
\label{eq_A}
\end{eqnarray}
where $u^{(l)}_{ij} = (\partial_i u_j^{(l)} + \partial_j u_i^{(l)})/2$ is  strain tensor,
$\gamma_0 = |V_{pp\pi}^0|$ is the nearest neighbor transfer energy of intrinsic graphene, and
\begin{eqnarray}
\beta = -\frac{d\ln T(\Vec{r}) }{d\ln r}\Bigr|_{r=a_0},
\end{eqnarray}
where $\Vec{r}$ is on the $xy$-plane.
In the present model Eq.\  (\ref{eq_transfer_integral}), we have $\beta = a_0/r_0 \approx 3.14$.
In the Fourier representation, Eq.\ (\ref{eq_A}) becomes
\begin{align}
\Vec{A}^{(l)}
=
\xi\frac{3}{4}\frac{\beta\gamma_0}{ev}
\sum_{\Vec{q}} 
\hat{W}_{\Vec{q}}
\Vec{u}_{\Vec{q}}^{(l)} e^{i\Vec{q}\cdot\Vec{r}},
\label{eq_delta_A}
\end{align}
where 
\begin{align}
\hat{W}_{\Vec{q}} = 
\begin{pmatrix}
i q_x  & -i q_y
\\
-i q_y & -i q_x 
\end{pmatrix}.
\label{eq_W}
\end{align}

\begin{figure}
\begin{center}
\leavevmode\includegraphics[width=0.9\hsize]{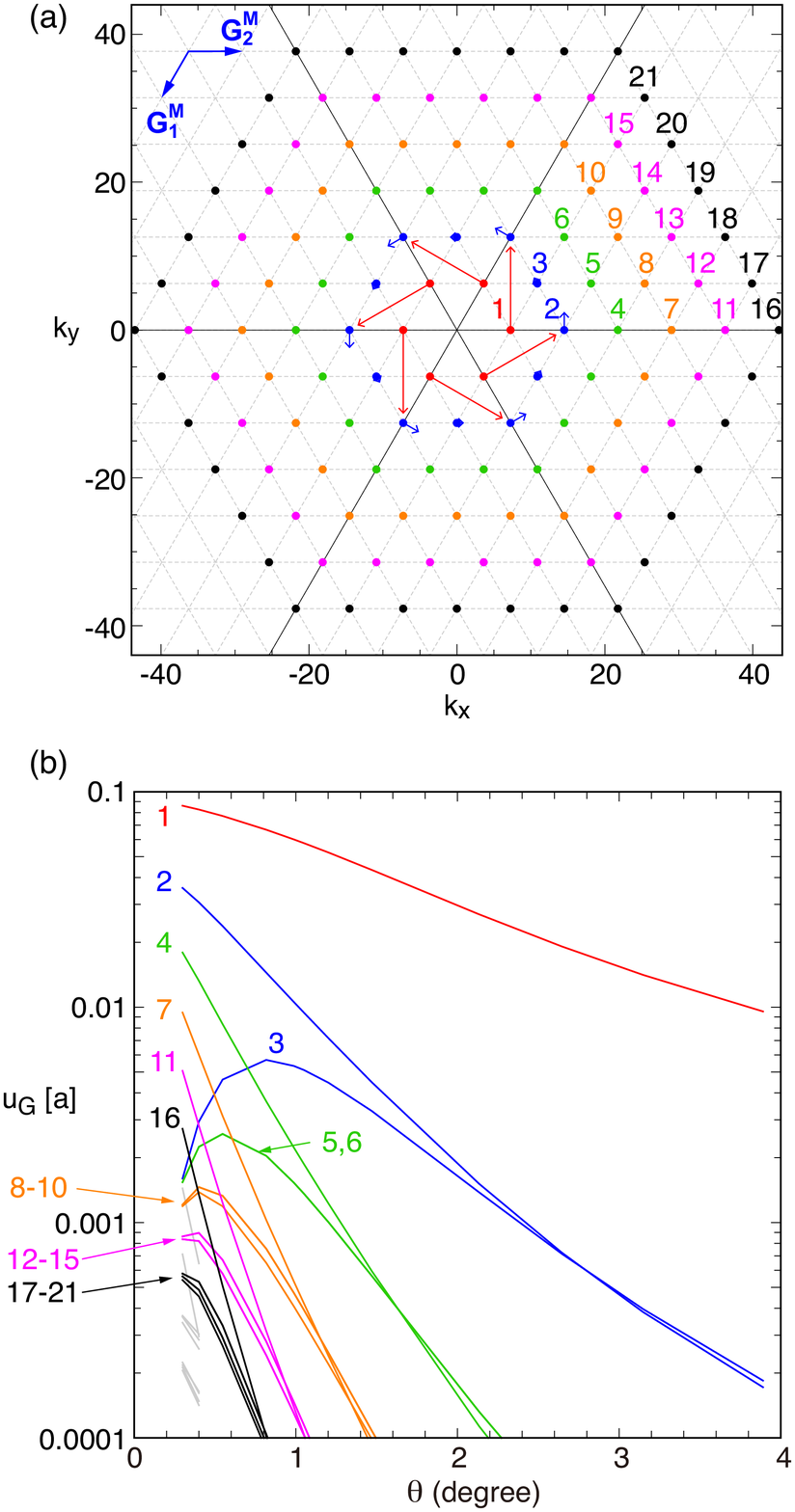}
\end{center}
\caption{
(a) The $k$-space map
of the Fourier components $i \textbf{u}^{-}_{\textbf{G}}$ at $\theta = 0.817^\circ$.
The grid points represents $\textbf{G}$'s and the length of arrows is proportional to $|\textbf{u}^{-}_{\textbf{G}}|$.
(b) Logarithmic plot of $|\textbf{u}^{-}_{\textbf{G}}|$ (in units of graphene's lattice constant, $a$)
as a function of the twist angle, where the numbers specify
the different $\Vec{G}$ vectors indicated in (a). Gray lines are for additional $\Vec{G}$ vectors taken only for $\theta =0.3^\circ$ and 0.4$^\circ$. 
}
\label{fig_displacement}
\end{figure}

\begin{figure*}
\begin{center}
\leavevmode\includegraphics[width=0.85\hsize]{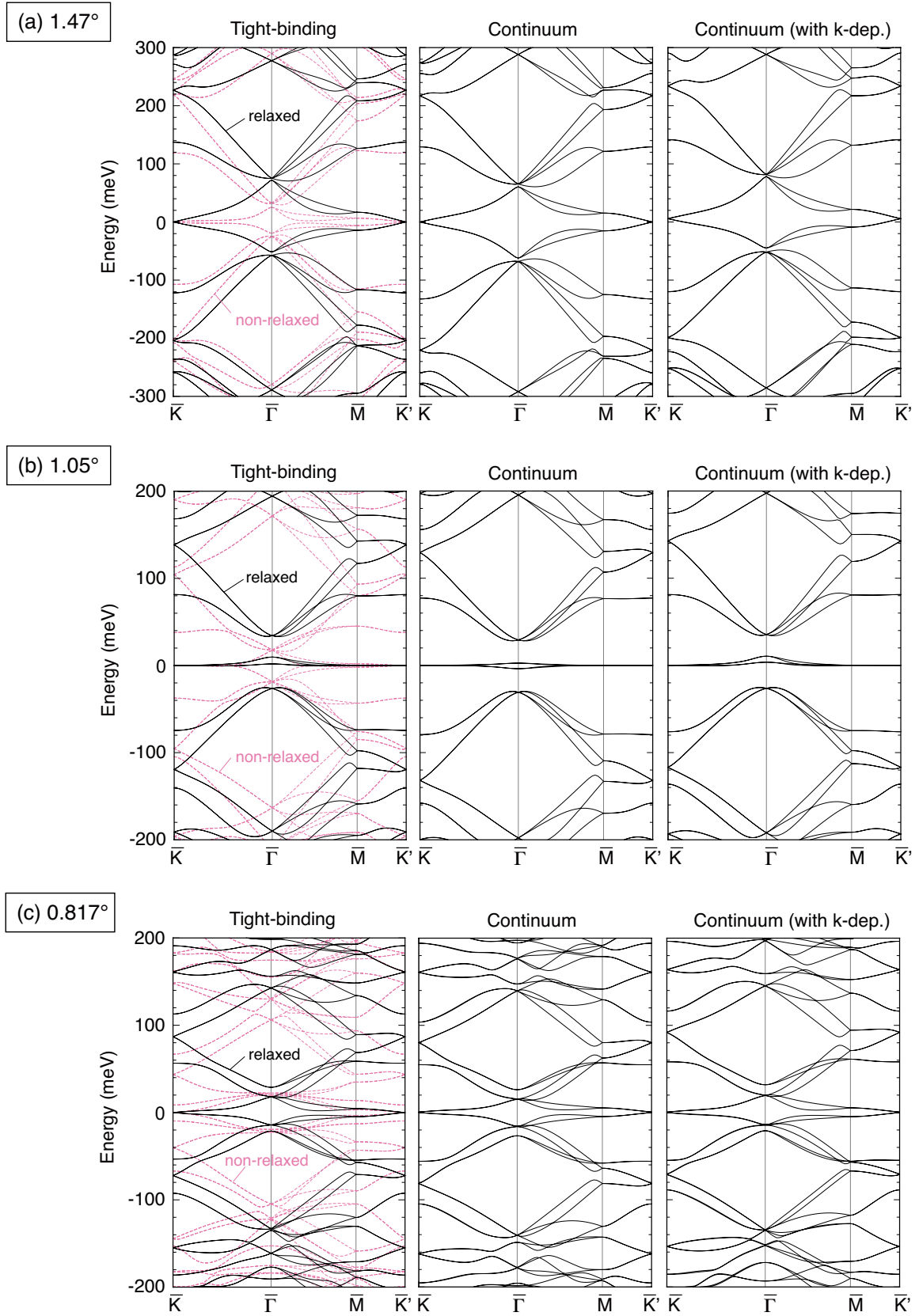}
\end{center}
\caption{
Band structure of relaxed TBGs of (a) $\theta = 1.47^\circ$, (b) $1.05^\circ$,
and (c) $0.817^\circ$.
The left panel is calculated by the original tight-binding model,
where dashed pink lines are the energy bands without lattice relaxation.
The middle and right panels are by the continuum model 
with the constant interlayer matrix $U$
and that with a linear $\Vec{k}$-dependence in $U$, respectively (see the text).}
\label{fig_band}
\end{figure*}

\section{Band structure of the relaxed TBG}
\label{sec_band}

Using the formula obtained above,
we calculate the band structure of relaxed TBGs with the AB-BA domain wall formation.
For the displacement vector, we use our previous calculation method \cite{nam2017lattice, koshino2019moire}
which considers only in-plane components, as the simplest approximation to describe the domain formation. 
Here $\Vec{u}^-(\Vec{r})$ is assumed to have the same periodicity as the original moir\'{e} pattern, or
\begin{align}
  & \textbf{u}^{-} (\textbf{r})  = 
\sum_{\textbf{G}} \textbf{u}^{-}_{\textbf{G}} e^{i\textbf{G}\cdot \textbf{r}},
\end{align}
where $\textbf{G} = m \textbf{G}^{\rm M}_1 + n \textbf{G}^{\rm M}_2$ are moir\'{e} reciprocal vectors.

Figure \ref{fig_displacement}(a) shows the $k$-space map
of the Fourier components $i \textbf{u}^{-}_{\textbf{G}}$ at $\theta = 0.817^\circ$,
where the triangular grid presents the moir\'{e} reciprocal lattice points
and the length of the arrows is proportional to $|\textbf{u}^{-}_{\textbf{G}}|$.
The $\textbf{u}^{-}$ has six-fold rotational symmetry as assumed in the calculation,
and its direction spirals around the origin.
Figure \ref{fig_displacement}(b) is a logarithmic plot of the absolute value $|\textbf{u}^{-}_{\textbf{G}}|$
as a function of the twist angle, where the numbers specify
the different $\Vec{G}$ vectors indicated in Fig.\ \ref{fig_displacement}(a).
In twist angles larger than $1^\circ$, the dominant contribution mostly comes from the shortest $\textbf{G}$
(indicated by  ``1'').
The higher harmonics becomes gradually relevant in $\theta < 1^\circ$,
where we see that the components at $\arctan (G_y/G_x) = 0^\circ, 60^\circ, 120^\circ,\cdots$
have relatively larger amplitudes than other directions.
We also found that, in any twist angles, $\textbf{u}^{-}_{\textbf{G}}$ and $\textbf{G}$ are exactly perpendicular
at the wave points with $\arctan (G_y/G_x) = 0^\circ, 30^\circ, 60^\circ,\cdots$,
and otherwise they are almost perpendicular with a few degree shift.
In the calculation, we took 21 $\Vec{G}$-points per 1/6 sector [i.e., $0 \leq \arctan (G_y/G_x) < 60^\circ$]
for twist angles $\theta > 0.5^\circ$, and 36 points for the two smallest angles, $\theta = 0.3^\circ$ and $0.4^\circ$.

Using the above $\textbf{u}^{-}$,
the Hamiltonian for the relaxed TBG is obtained by using Eqs.\ (\ref{eq_umat_in-plane}) and (\ref{eq_H_eff}).
In the practical calculation, we expand $\exp[i \Vec{Q}_j\cdot \Vec{u}^-(\Vec{r})]$ in Eq.\ (\ref{eq_umat_in-plane})
back to the $k$-space representation Eq.\ (\ref{eq_umat_3d_approx1}), and diagonalize the Hamiltonian in $k$-space bases.
The matrix element is expressed in power of $|\Vec{Q}_j\cdot\Vec{u}^-_\Vec{G}|$
as in Eq.\ (\ref{eq_gamma_func_in-plane}), and the higher harmonics quickly decay provided that
$|\Vec{Q}_j\cdot\Vec{u}^-_\Vec{G}|$ is much smaller than 1.
This condition is well satisfied in the twist angles studied here,
since $|\Vec{Q}_j| = 4\pi/(3a)$ and $|\Vec{u}^-_\Vec{G}| < 0.1a$ as seen in Fig.\ \ref{fig_displacement}(b).
Figure \ref{fig_band} presents the calculated band structures of relaxed TBGs at twist angles
(a) 1.47$^\circ$, (b) 1.05$^\circ$, and (c) 0.817$^\circ$.
In each row,
the left panel is the band structure calculated by the original tight-binding model 
with the hopping function of Eq.\ (\ref{eq_transfer_integral}),
where solid black and dashed pink lines are the energy bands with and without lattice relaxation,
respectively, and we shifted the origin of the energy axis to the band touching points at $\bar{K}$ and $\bar{K}'$.
The middle panel presents the energy band of relaxed TBGs obtained by the 
continuum model of Eqs.\ (\ref{eq_umat_in-plane}) and (\ref{eq_H_eff}).

We see a nice agreement between the tight-binding model and the continuum model,
while also notice that a slight asymmetry between the electron side and hole side 
in the tight-binding result is ignored in the continuum model.
Actually, it was recently shown that the electron-hole asymmetry can be taken into account 
by including $k$-linear term in the interlayer coupling. \cite{carr2019exact,fang2019angle,guinea2019continuum}
Our original expression for the interlayer matrix element
Eq.\ (\ref{eq_umat_3d}) depends on the position of initial $\Vec{k}$
through $\Vec{Q}$, but the $\Vec{k}$-dependence is dropped 
by replacing $\Vec{Q}$ with constant $\Vec{Q}_j$ in Eq.\ (\ref{eq_umat_in-plane}).
In the right panels, we show the band structure calculated by the original Eq.\ (\ref{eq_umat_3d}) 
with $\Vec{k}$-dependence included in the linear order,
where we actually see that the electron-hole asymmetry is restored.

The role of the displacement vector can be understood by 
expanding the Hamiltonian in powers of the displacement vectors.
Within the first order in $\Vec{u}^{-}$, the interlayer matrix $U$ of Eq.\ (\ref{eq_umat_in-plane})
is written as
\begin{align}
U_{X'X} (\Vec{r}) \approx
t_0 \sum_{j=1}^3
M^j_{X'X} e^{i \delta\Vec{k}_j\cdot\Vec{r}}
\Biggl(1+i \Vec{Q}_j\cdot 
\sum_{\textbf{G}} \textbf{u}^{-}_{\textbf{G}} e^{i\textbf{G}\cdot \textbf{r}}
\Biggr),
\label{eq_umat_in-plane_linear}
\end{align}
By only taking the six dominant components of $\Vec{u}^{-}_\Vec{G}$ 
[Fig.\ \ref{fig_displacement}(a)],
we have
\begin{align}
 U &= 
\begin{pmatrix}
U_{A A} & U_{A B}
\\
U_{B A} & U_{B B}
\end{pmatrix}
\nonumber\\
&
\approx
\begin{pmatrix}
t & t'
\\
t' & t
\end{pmatrix}
+
\begin{pmatrix}
t & t'\omega^{-\xi}
\\
t'\omega^\xi & t
\end{pmatrix}
e^{i\xi \Vec{G}^{\rm M}_1\cdot\Vec{r}}
\nonumber\\
& 
 \qquad +
\begin{pmatrix}
t & t'\omega^\xi
\\
t'\omega^{-\xi} & t
\end{pmatrix}
e^{i\xi(\Vec{G}^{\rm M}_1+\Vec{G}^{\rm M}_2)\cdot\Vec{r}}
+ \cdots,
\label{eq_umat_approx}
\end{align}
where
\begin{align}
&t = t_0 (1-2\alpha), \quad t' = t_0 (1+\alpha/2),
\\
& \alpha = \frac{2\pi}{\sqrt{3}}\frac{u_1}{a},
\end{align}
and $u_1$ is the absolute value of the leading $\Vec{u}^{-}_{\Vec{G}}$,
plotted as curve 1 in Fig.\ \ref{fig_displacement}(b).
We see that the in-plane distortion enhances the off-diagonal elements (AB and BA) 
while suppresses the diagonal elements (AA and BB) in the $U$ matrix.
This is interpreted  as a consequence of the lattice relaxation,
which maximizes the AB/BA-stacking area while minimizes the unfavorable
AA/BB-stacking area.

Interestingly, a similar Hamiltonian with $t < t'$ was also obtained 
by considering the out-of-plane distortion only.\cite{koshino2018maximally}
In this case, the diagonal terms is reduced because 
the interlayer spacing at $AA$ region is elongated and the local interlayer coupling is reduced.
In the band structure, the difference between $t$ and $t'$ is responsible for the gap opening between
the flat band and the excited bands\cite{koshino2018maximally,tarnopolsky2019origin,liu2019pseudo},
and this is also true in the present case with the in-plane distortion, as shown in Fig.\ \ref{fig_band}.

\section{Electron-phonon interaction in the relaxed TBG}
\label{sec_e-p}

\subsection{Quantized moir\'{e} phonons in TBG}
  
Here we derive the Hamiltonian of moir\'{e} acoustic phonons in the relaxed TBG,
by quantizing the classical motion of the lattice vibration. \cite{koshino2019moire}
The interaction between the quantized phonons and the electronic system
will be argued in the next subsection.

We consider a long-wave, in-plane lattice vibration specified by the displacement vector,
$\textbf{u}^{(l)}_A = \textbf{u}^{(l)}_B = \textbf{u}^{(l)}(\textbf{r},t)$ 
for layer $l =1,2$. We again assume $u_z^{(l)} = 0$.
The Lagrangian of the system is given by $L=T-(U_E+U_B)$
as a functional of $\textbf{u}^{(l)}$.
The term $T$ is the kinetic energy due to the motion of the carbon atoms,
\begin{align}
 T= \sum_{l=1}^2 \int 
 \frac{\rho}{2} \left[\dot{u}_{x}^{(l)2} + \dot{u}_{y}^{(l)2}\right] d^2 r,
 \label{Eq:T_TBG}
\end{align}
where $\rho= 7.61\times 10^{-7}$ kg/m$^2$ is the area density of single-layer graphene,
and $\dot{u_i}$ represents the time derivative of $u_i$.
The $U_E$  is the elastic energy of strained TBG given by \cite{suzuura2002phonons,San-Jose2014GhBN}
\begin{align}
 U_E = \sum_{l=1}^2 \int \frac{1}{2} \left\{ (\lambda +\mu ) (u_{xx}^{(l)} + u_{yy}^{(l)})^2 \right. \nonumber\\
 \left.   +\mu \left[ ( u_{xx}^{(l)} - u_{yy}^{(l)})^2 + 4(u_{xy}^{(l)})^2 \right] \right\} d^2 r,
 \label{Eq:U_E_TBG}
\end{align}
 where 
 $\lambda \approx 3.25$ eV/$\text{\AA}^2$ and
 $\mu \approx 9.57$ eV/$\text{\AA}^2$ are graphene's 
 $\text{Lam}\acute{\text{e}}$  factors \cite{zakharchenko2009finite,jung2015origin}, 
and $u^{(l)}_{ij} = (\partial_i u_j^{(l)} + \partial_j u_i^{(l)})/2$ is  strain tensor.
The $U_B$ is the registry-dependent inter-layer binding energy  \cite{nam2017lattice, koshino2019moire},
\begin{align}
\label{Eq:U_B_TBG}
U_B  = \int \text{d}^2 r \,
\sum^3_{j= 1}  
2V_0 \cos [ \textbf{G}_j^{\rm M} \cdot \textbf{r} + \textbf{a}^*_j\cdot(\textbf{u}^{(2)} -\textbf{u}^{(1)}) ],
\end{align}
where $\textbf{a}^*_3 = -\textbf{a}^*_1 - \textbf{a}^*_2$,
and $\textbf{G}^{\rm M}_3 = - \textbf{G}^{\rm M}_1 - \textbf{G}^{\rm M}_2$.
The difference between the binding energies of AA and AB/BA structure 
is $9V_0$ per area, and this amounts to $\Delta  \epsilon = 9V_0 S_G/ 4$ per atom
where $S_G$ is the area of graphene's unit cell.
In the following calculation, we use $\Delta \epsilon = 0.0189$ (eV/atom) as a typical value \cite{Lebedeva2011,popov2011commensurate}.

The Lagrangian $L$ can be separated into the interlayer symmetric part and asymmetric part,
which are associated with $\textbf{u}^{\pm} = \textbf{u}^{(2)} \pm \textbf{u}^{(1)}$, respectively.
Since the interlayer binding energy $U_B$ only depends on $\textbf{u}^{-}$,
the moir\'{e} interlayer coupling only affects the motion of $\textbf{u}^{-}$
while leaving $\textbf{u}^{+}$ unchanged from the intrinsic graphene.\cite{koshino2019moire}
In the following, we only consider $\textbf{u}^{-}$ sector of the Lagrangian.
We consider a small vibration around the relaxed state, i.e.,
\begin{align}
\textbf{u}^{-} (\Vec{r},t) =  \textbf{u}^{-}_0 (\Vec{r}) + \delta\textbf{u}^{-} (\Vec{r},t).
\label{eq_u_plus_du}
\end{align}
Here $\textbf{u}^{-}_0 (\Vec{r})$ is the static relaxed state to minimize $U_B+U_E$,
which was argued in Sec.\ \ref{sec_band},
and $\delta\textbf{u}^{-} (\Vec{r},t)$ is a perturbational vibration around $\textbf{u}^{-}_0$.
We define the Fourier transform 
 \begin{align}
 & \textbf{u}_0^{-} (\textbf{r})  = 
\sum_{\textbf{G}} \textbf{u}^{-}_{0,\textbf{G}} e^{i\textbf{G}\cdot \textbf{r}},
\label{eq_u0_fourier}
\\
& \delta\textbf{u}^{-} (\textbf{r},t)  = 
\sqrt{\frac{1}{S}}
\sum_{\textbf{q}}\delta\textbf{u}^{-}_{\textbf{q}}(t)  e^{i\textbf{q}\cdot \textbf{r}},
\label{eq_delta_u_fourier}
\end{align}
where $\Vec{G} = m_1 \Vec{G}^M_1 + m_2 \Vec{G}^M_2$,
and the factor $\sqrt{1/S}$ is required to normalize the phonon operators introduced later.
   
We rewrite the Lagrangian in terms of  $\textbf{u}_0^{-}$ and $\delta \textbf{u}^{-}$,
and expand it into a series of  $\delta \textbf{u}^{-}$ within the second order.
The relaxed state $\textbf{u}^{-}_0$ can be obtained by the variational principle
$[\delta (U_E+U_B)/ \delta (\delta \textbf{u}^{-}) ] = 0$. \cite{nam2017lattice}
We introduce the canonical momentum
\begin{align}
\delta \Vec{p}^-_\Vec{q} = \frac{\partial L}{\partial \delta \textbf{u}^{-}_{\Vec{q}}}
= \rho_r \, \delta \dot{\textbf{u}}^{-}_{-\Vec{q}},
\end{align}
where 
\begin{align}
\rho_r = \frac{\rho}{2}
\end{align}
is the reduced mass for the relative motion.
The Hamiltonian 
$H = \sum_\Vec{q}  \delta \Vec{p}^-_\Vec{q} \cdot \delta \textbf{u}^{-}_{\Vec{q}} - L$
can be written as
\begin{align}
H = 
&\sum_{\Vec{q}\in{\rm MBZ}}
\Biggl[
\sum_\Vec{G}
\frac{1}{2\rho_r}
\delta \Vec{p}^{-\dagger}_\Vec{q+\Vec{G}} \cdot \delta\Vec{p}^-_\Vec{q+\Vec{G}} 
\nonumber
\\
&\quad + \sum_{\Vec{G},\Vec{G}'}
\frac{1}{2} 
\delta\textbf{u}^{-\dagger}_{\Vec{q}+\Vec{G}'} 
\hat{D}_{\Vec{q}}(\Vec{G}',\Vec{G})
\delta\textbf{u}^{-}_{\Vec{q}+\Vec{G}} 
\Biggr],
\label{eq_H_phonon}
\end{align}
where $\Vec{G}$ and  $\Vec{G}'$  run over the moir\'{e} reciprocal lattice vectors 
$m_1 \Vec{G}^M_1 + m_2 \Vec{G}^M_2$,
and MBZ represents the first moir\'{e} Brillouin zone spanned by $\Vec{G}^M_1$ and $\Vec{G}^M_2$.
Here we use $\sum_\Vec{q} f_\Vec{q} = \sum_{\Vec{q}\in{\rm MBZ}}\sum_\Vec{G} f_{\Vec{q}+\Vec{G}}$
for arbitrary function $f_\Vec{q}$.
The $\hat{D}$ is the dynamical matrix given by
\begin{align} 
&       \hat{D}_{\Vec{q}}(\Vec{G}',\Vec{G}) =
\frac{1}{2}  \hat{K}_{\Vec{q}+\textbf{G}} \delta_{\Vec{G},\Vec{G}'}
+ \hat{V}_{\textbf{G}'-\textbf{G}},
\end{align}
where
\begin{align}
&        \hat{K}_{\textbf{q}}=\left( {\begin{array}{cc}
        (\lambda+2\mu) q_x^2 + \mu q_y^2 &  (\lambda + \mu)q_x q_y \\
        (\lambda +\mu) q_x q_y & (\lambda + 2\mu)q_y^2 + \mu q_x^2
        \end{array}} \right),
 \nonumber
 \\
&        \hat{V}_{\textbf{G}}=
(-2V_0)  \sum_{j =1}^3  h^j_{\textbf{G}} 
\begin{pmatrix}
a^*_{jx}a^*_{jx} & a^*_{jx}a^*_{jy}\\
a^*_{jx}a^*_{jy} & a^*_{jy}a^*_{jy}
\end{pmatrix},
\end{align}
Here $a^*_{j\mu}$ is the $\mu$ component of $\Vec{a}_j^*$, and $h^j_{\textbf{G}}$ is defined by
 \begin{align}
& \cos \left[\textbf{G}_j^{\rm M} \cdot \textbf{r} + \textbf{a}^*_j \cdot \textbf{u}^{-}_0(\textbf{r})\right] 
= \sum_{\textbf{G}} h^j_{\textbf{G}} e^{i \textbf{G} \cdot \textbf{r}}.
\end{align}
 
 For each $\Vec{q}$ in MBZ, the eigen modes can be found by the secular equation,
 \begin{align} 
 \sum_{\textbf{G}'} \hat{D}_{\Vec{q}}(\textbf{G},\textbf{G}') \textbf{C}_{n,\textbf{q}}(\Vec{G}')
= \rho_r \, \omega_{n,\Vec{q}}^2
\textbf{C}_{n,\textbf{q}}(\Vec{G}),
\end{align}
where $n$ is the mode index, $\omega_{n,\Vec{q}}$ is the eigen frequency,
and $\textbf{C}_{n,\textbf{q}}(\Vec{G}) = (C^x_{n,\textbf{q}}(\Vec{G}), C^y_{n,\textbf{q}}(\Vec{G}))$
is the eigenvector normalized by $\sum_\Vec{G} |\textbf{C}_{n,\textbf{q}}(\Vec{G})|^2 = 1 $. 
By applying a unitary transformation,
\begin{align}
&\delta\Vec{u}^-_\Vec{q+\Vec{G}}  
= \sum_n \textbf{C}_{n,\textbf{q}}(\Vec{G}) \delta \tilde{u}_{n,\Vec{q}},
\nonumber\\
&\delta\Vec{p}^-_\Vec{q+\Vec{G}} 
= \sum_n \textbf{C}^*_{n,\textbf{q}}(\Vec{G}) \delta \tilde{p}_{n,\Vec{q}},
\label{eq_u_unitary}
\end{align}
the Hamiltonian Eq.\ (\ref{eq_H_phonon}) is written as a diagonal form
\begin{align}
H = 
&\sum_{\Vec{q}\in{\rm MBZ}}
\sum_n
\frac{1}{2\rho_r}
\delta \tilde{p}^\dagger_{n,\Vec{q}}  \delta \tilde{p}_{n,\Vec{q}}
+ \frac{\rho_r}{2} 
\omega_{n,\Vec{q}}^2
\delta \tilde{u}^\dagger_{n,\Vec{q}}  \delta \tilde{u}_{n,\Vec{q}}.
\label{eq_H_phonon2}
\end{align}
We introduce the canonical quantization by
$[ \delta \tilde{u}_{n,\Vec{q}},  \delta \tilde{p}_{n',\Vec{q}'}] = i \hbar \delta_{n,n'}\delta_{\Vec{q},\Vec{q}'}$.
We define the phonon creation and annihilation operators $a^\dagger_{n,\Vec{q}}, a_{n,\Vec{q}}$ by
\begin{align}
& \delta \tilde{u}_{n,\Vec{q}}  = \sqrt{\frac{\hbar}{2\rho_r \omega_{n,\Vec{q}}}} 
(a_{n,\Vec{q}} + a^\dagger_{n,-\Vec{q}})
\nonumber\\
& \delta \tilde{p}_{n,\Vec{q}}  = i \sqrt{\frac{\rho_r \hbar \omega_{n,\Vec{q}}}{2}} 
(a^\dagger_{n,\Vec{q}} - a_{n,-\Vec{q}}),
\label{eq_phonon_operators}
\end{align}
which satisfies $[a_{n,\Vec{q}},a^\dagger_{n',\Vec{q}'}]=\delta_{n,n'}\delta_{\Vec{q},\Vec{q}'}$.
Finally, the Hamiltonian becomes
\begin{align}
H = 
&\sum_{\Vec{q}\in{\rm MBZ}}
\sum_n
\hbar \omega_{n,\Vec{q}}
\left(
a^\dagger_{n,\Vec{q}}  a_{n,\Vec{q}} + \frac{1}{2}
\right).
\label{eq_H_phonon3}
\end{align}

\subsection{Electron-phonon matrix elements}

The electron phonon interaction is contributed by the interlayer part
and the intralayer part, where the former 
originates from the change of the moir\'{e} pattern
and the latter from the strain-induced pseudo vector field.
The interlayer part is obtained by replacing Eq.\ (\ref{eq_umat_in-plane})
with $\textbf{u}^{-}  =  \textbf{u}^{-}_0 + \delta\textbf{u}^{-}$ and taking 
the first order in $\delta\textbf{u}^{-}$.
As we consider the long-range phonons here,
the electron phonon scattering occurs only within a single valley $\xi$.

The change in the interlayer Hamiltonian Eq.\ (\ref{eq_umat_in-plane}) is 
written as
\begin{align}
&\delta U_{X'X} (\Vec{r}) 
= t_0 \sum_{j=1}^3
M^j_{X'X} \exp[i \Vec{Q}_j\cdot \Vec{u}_0^-(\Vec{r}) + i \delta\Vec{k}_j\cdot\Vec{r}] 
\nonumber\\
&\qquad\qquad\qquad
\times[i \Vec{Q}_j\cdot \delta\Vec{u}^-(\Vec{r},t)],
\end{align}
By using Eqs. (\ref{eq_delta_u_fourier}), (\ref{eq_u_unitary}) and (\ref{eq_phonon_operators}),
$\delta\Vec{u}^-$ can be expressed in terms of the phonon operators as,
\begin{align}
& \delta\textbf{u}^{-} (\textbf{r},t)  = 
\sqrt{\frac{1}{S}}
\sum_{\Vec{q}\in{\rm MBZ}}
\sum_\Vec{G}
\delta\textbf{u}^{-}_{\textbf{q}+\Vec{G}}  e^{i(\textbf{q}+\Vec{G})\cdot \textbf{r}}
\nonumber\\
&
 =
\sum_{n,\Vec{q}\in{\rm MBZ}}
\sum_\Vec{G}
\textbf{C}_{n,\textbf{q}}(\Vec{G}) \, e^{i(\textbf{q}+\Vec{G})\cdot \textbf{r}}
\nonumber\\
&
\qquad\qquad
\times \sqrt{\frac{1}{S}\frac{\hbar}{2\rho_r \omega_{n,\Vec{q}}}} 
(a_{n,\Vec{q}} + a^\dagger_{n,-\Vec{q}}).
\end{align}
Finally, the matrix element for the interlayer part of electron-phonon coupling is written as
\begin{align}
& \langle \Vec{k}',X',2 |H_{\mbox{\scriptsize e-p}}| \Vec{k},X,1 \rangle
= \frac{1}{S} \int d^2 r
e^{i(\Vec{k}-\Vec{k}')\cdot\Vec{r}}
\delta U_{X'X} (\Vec{r})
\nonumber\\
&
= \frac{1}{\sqrt{S}}
\sum_{n,\Vec{q}\in{\rm MBZ}} (a_{n,\Vec{q}} + a^\dagger_{n,-\Vec{q}}) 
\, g_{n,\Vec{q}} (\Vec{k}',X',2; \Vec{k},X,1),
\end{align}
where the electron-phonon coupling strength $g_{n,\Vec{q}}$ is given by,
\begin{align}
&g_{n,\Vec{q}} (\Vec{k}',X',2; \Vec{k},X,1) 
\nonumber\\
&= 
t_0 
\sqrt{\frac{\hbar}{2\rho_r \omega_{n,\Vec{q}}}}
\sum_{j=1}^3
M^j_{X'X} 
\sum_\Vec{G}
i\Vec{Q}_j \cdot \textbf{C}_{n,\textbf{q}}(\Vec{G}) 
\nonumber\\
&\qquad
\times
\frac{1}{S} \int d^2 r\,
e^{i(\Vec{k} -\Vec{k}' + \delta\Vec{k}_j + \textbf{q}+\Vec{G})\cdot\Vec{r}}
e^{i \Vec{Q}_j\cdot \Vec{u}_0^-(\Vec{r})}.
\label{eq_g_inter}
\end{align}


On the other hand, the change in the intralayer Hamiltonian Eq.\ (\ref{eq_intra}) is 
\begin{align}
& \delta H_l = - e v \left[
R(\pm \theta/2)
\, \delta \Vec{A}^{(l)}
\right]
 \cdot (\xi \sigma_x, \sigma_y),
\label{eq_intra_delta}
\end{align}
where $\mp$ is for $l=1$ and 2, respectively, and 
$\delta \Vec{A}^{(l)}$ is the shift of the pseudo vector  field Eq.\ (\ref{eq_A}), or
\begin{align}
\delta \Vec{A}^{(l)}
=
\xi\frac{3}{4}\frac{\beta\gamma_0}{ev}
\sqrt{\frac{1}{S}}
\sum_{\Vec{q}} 
\hat{W}_{\Vec{q}}
\delta \Vec{u}_{\Vec{q}}^{(l)} e^{i\Vec{q}\cdot\Vec{r}},
\label{eq_delta_A}
\end{align}
where $\hat{W}_{\Vec{q}}$ is defined in Eq.\ (\ref{eq_W}).
When we consider the interlayer asymmetric modes,
we have $\delta \Vec{u}_{\Vec{q}}^{(l)} = \mp \delta \Vec{u}_{\Vec{q}}^{-}/2$ with $\mp$ for $l=1$ and 2, respectively.
We again use Eqs.\ (\ref{eq_u_unitary}) and (\ref{eq_phonon_operators})
to write $\delta \Vec{A}^{(l)}$ in terms of the phonon operators.
The intralayer part of electron-phonon coupling is finially written as
\begin{align}
& \langle \Vec{k}',X',l |H_{\mbox{\scriptsize e-p}}| \Vec{k},X,l \rangle
\equiv \frac{1}{S} \int d^2 r
e^{i(\Vec{k}-\Vec{k}')\cdot\Vec{r}}
(\delta H_l)_{X'X} 
\nonumber\\
\nonumber\\
&
=  \frac{1}{\sqrt{S}} \sum_{n,\Vec{q}\in{\rm MBZ}} (a^\dagger_{n,\Vec{q}} + a_{n,-\Vec{q}}) 
\, g_{n,\Vec{q}} (\Vec{k}',X',l; \Vec{k},X,l),
\end{align}
where
\begin{align}
&g_{n,\Vec{q}} (\Vec{k}',X',l; \Vec{k},X,l) 
= 
\pm \xi \frac{3}{8} \beta\gamma_0 
\sqrt{\frac{\hbar}{2\rho_r \omega_{n,\Vec{q}}}}
\nonumber\\
&\quad \times
\sum_\Vec{G}
\left[
R(\pm \theta/2)\hat{W}_{\Vec{q}+\Vec{G}}
 \textbf{C}_{n,\textbf{q}}(\Vec{G}) 
\cdot (\xi \sigma_x, \sigma_y)
\right]_{X',X}
\nonumber\\
&\quad \times
\delta_{\Vec{k}+\textbf{q}+\Vec{G},  \Vec{k}'},
\label{eq_g_intra}
\end{align}
where $\pm$ is for $l=1$ and 2, respectively.

\begin{figure*}
\begin{center}
\leavevmode\includegraphics[width=1.\hsize]{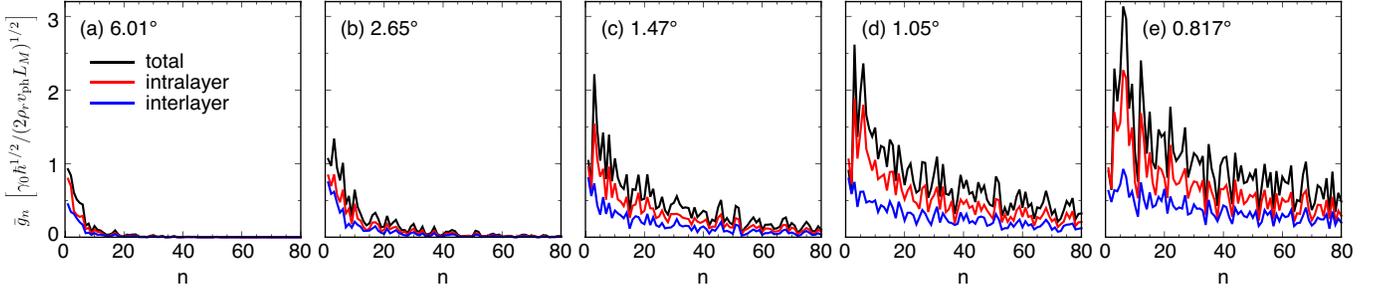}
\end{center}
\caption{
Plots of $\bar{g}_{n}$ as a function of the phonon band index $n$
in different twist angles. Red and blue curves are the intralayer [Eq.\ (\ref{eq_g_intra})] and 
interlayer [Eq.\ (\ref{eq_g_inter})] contributions, respectively, and the black curve is the total amplitude.
}
\label{fig_gn}
\end{figure*}

In the following, we numerically calculate the electron-phonon coupling for the lowest bands in TBG.
The eigenstates of TBG is written as
\begin{align}
|\alpha, \Vec{k}\rangle 
=
\sum_{\Vec{G} X l}
F_{\alpha,\Vec{k}}(\Vec{G},X,l)
\,| \Vec{k}+\Vec{G},X,l \rangle,
\end{align}
where $\alpha$ is the band index and $\Vec{k}$ is the Bloch vector in MBZ.
The electron-phonon coupling is expressed in the eigenstate basis as
\begin{align}
&H_{\mbox{\scriptsize e-p}} =
\frac{1}{\sqrt{S}} 
 \sum_{n \alpha \alpha'} \sum_{\Vec{q} \Vec{k} \Vec{k}'\in{\rm MBZ}} 
  (a^\dagger_{n,\Vec{q}} + a_{n,-\Vec{q}}) 
c^\dagger_{\alpha',\Vec{k}'} c_{\alpha,\Vec{k}} \, 
 \nonumber\\
& \hspace{40mm} \times g_{n,\Vec{q}} (\alpha', \Vec{k}'; \alpha, \Vec{k}),
\end{align}
where $c^\dagger_{\alpha, \Vec{k}}$ and $c_{\alpha, \Vec{k}}$ are creation and annihilation operators, respectively,
of an electron in the state $|\alpha, \Vec{k}\rangle$, and we defined
\begin{align}
& g_{n,\Vec{q}} (\alpha', \Vec{k}'; \alpha, \Vec{k}) =
\nonumber \\
& \hspace{10mm} 
 \sum_{\Vec{G} X l}\sum_{\Vec{G}' X' l'}
g_{n,\Vec{q}} (\Vec{k}'+\Vec{G}',X',l; \Vec{k}+\Vec{G},X,l) 
\nonumber \\
& \hspace{25mm} \times F^*_{\alpha',\Vec{k}'}(\Vec{G}',X',l')F_{\alpha,\Vec{k}}(\Vec{G},X,l).
\end{align}
The coupling strength $g_{n,\Vec{q}} (\alpha', \Vec{k}'; \alpha, \Vec{k})$ becomes non-zero 
only when $ \Vec{k}' = \Vec{k}+\Vec{q}+\Vec{G}$ with moir\'{e} lattice vector $\Vec{G}$.

To characterize the electron-phonon coupling strength in the low-lying bands,
we define the averaged coupling amplitude $\bar{g}_{n}$ as
\begin{align}
\bar{g}_{n}^2 \equiv
\frac{1}{N_\Vec{q}} \sum_{\Vec{q}\in{\rm MBZ}} \frac{1}{4} \sum_{\alpha=\pm}\sum_{\alpha'=\pm}
|g_{n,\Vec{q}} (\alpha', \bar{\Vec{K}}+\Vec{q}; \alpha, \bar{\Vec{K}})|^2,
\label{eq_g_n_bar}
\end{align}
where $\bar{\Vec{K}}$ is the position of the moir\'{e} Brillouin zone corner $\bar{K}$,
$N_\Vec{q}$ is the number of sampling points of $\Vec{q}$ in the MBZ, which is taken as 27 in this work.
The band indexes $\alpha = \pm$ represent the lowest electron band and hole band, respectively, in a single valley and spin sector,
which correspond to the nearly-flat bands at the magic angle TBG.
The factor $1/4$ averages the four different processes from $\alpha = \pm$ to $\alpha' = \pm$.  
Here we take $\bar{\Vec{K}}$ as the reference point, while the quantitative behavior does not depend on its choice.

Figure \ref{fig_gn} presents the plots of $\bar{g}_{n}$ as a function of the phonon band index $n$
calculated for different twist angles. The red and blue curves are the intralayer [Eq.\ (\ref{eq_g_intra})] and 
interlayer [Eq.\ (\ref{eq_g_inter})] contributions, respectively, and the black curve is the total amplitude.
Here the unit of the vertical axis is taken as  $\gamma_0 \sqrt{\hbar /(2\rho_r v_{\rm ph}L_M)}$
where $L_M$ is the size of the moir\'{e} unit cell
and $v_{\rm ph}$ is the typical phonon velocity of monolayer graphene.
We take $v_{\rm ph}=\sqrt{\mu/\rho} \approx 1.4\times10^4$ m/s, which is the velocity of the transverse acoustic phonon modes.

In large twist angles ($\theta \gsim 5^\circ$), we see that $\bar{g}_{n}$ decays quickly in $n$.
There the superlattice hybridization is week in the low-energy region, so that
each of electron and phonon eigenstates is  dominated by a monolayer eigenstate with a single wave component.
Then the phonons of large $n$'s (mainly composed of high $q$) do not have relevant matrix elements
in the low-lying electronic states (composed of low $k$'s), and this is the reason for the quick decay of $\bar{g}_{n}$.
The electron-phonon coupling in this regime is approximately given by
\begin{align}
g_q \sim \gamma_0 \sqrt{\frac{\hbar q}{2\rho_r v_{\rm ph}}},
\label{eq_gq_low_angle}
\end{align}
which is obtained by using $\omega_{n,\Vec{q}} \sim v_{\rm ph} q$ and 
$\hat{W}_{\Vec{q}+\Vec{G}} \sim q$ in Eq.\ (\ref{eq_g_intra}).
In the calculation of $\bar{g}_{n}$, the wavenumber $q$ is averaged in MBZ (of the size $\sim 1/L_M$), 
so the magnitude of $\bar{g}_n$ is roughly given by $\gamma_0 \sqrt{\hbar /(2\rho_r v_{\rm ph}L_M)}$,
which is the vertical unit in Fig. \ref{fig_gn}. 

In low twist angles, on the other hand, the wave functions spread over different $\Vec{G}$'s in the momentum space
due to the moir\'{e} superlattice hybridization, and then the phonon modes in large $n$ are able to couple the low-lying electronic states.
This is observed as a long tail in Fig. \ref{fig_gn}. 
In this regime, the typical order of magnitude of $\hat{W}_{\Vec{q}+\Vec{G}}$ is given by
the momentum-space distribution range of the electronic states,  which is of the order of
\begin{align}
\Delta q \sim \frac{t_0}{\hbar v}.
\end{align}
The phonon frequency $\omega_{n,\Vec{q}}$ is of the order of $v_{ph} /L_M$, 
considering the band folding of the linear phonon dispersion. \cite{koshino2019moire}
As a result, the characteristic magnitude of the intralayer electron-phonon coupling, Eq.\ (\ref{eq_g_intra}),
 becomes  
 \begin{align}
g^{\rm(intra)}_{n,\Vec{q}} \sim \gamma_0 \sqrt{\frac{\hbar L_M}{2\rho_r v_{\rm ph}}} \Delta q.
\label{eq_g_intra_order}
\end{align}
Since $L_M \propto \theta^{-1}$, the overall amplitude of $g^{\rm(intra)}_{n,\Vec{q}} $ 
increases in decreasing the twist angle $\theta$,
and this is actually observed in Fig.\ \ref{fig_gn}. 

The magnitude of the interlayer electron-phonon coupling Eq.\ (\ref{eq_g_inter})
is estimated as 
 \begin{align}
g^{\rm(inter)}_{n,\Vec{q}} \sim 
t_0 \sqrt{\frac{\hbar L_M}{2\rho_r v_{\rm ph}}} \frac{1}{a},
\label{eq_g_inter_order}
\end{align}
where we noted that $\Vec{Q}_j$ is of the order of $1/a$.
The relative magnitude of interlayer part to the intralayer part is 
\begin{align}
\frac{g^{\rm(inter)}_{n,\Vec{q}}} {g^{\rm (intra)}_{n,\Vec{q}}}
\sim \frac{t_0 (1/a)}{\gamma_0 \Delta q}
\sim \frac{1}{\gamma_0} \frac{\hbar v}{a}\approx 0.8.
\end{align}
In Fig. \ref{fig_gn}, we  actually see that the two components have comparable magnitudes,
while the interlayer contribution is always smaller about by a factor $\sim 2$.

\begin{figure}
\begin{center}
\leavevmode\includegraphics[width=0.85\hsize]{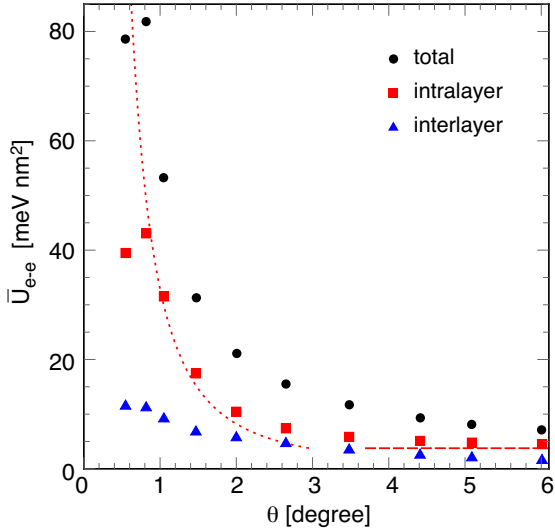}
\end{center}
\caption{
$\bar{U}_{\mbox{\scriptsize e-e}}$ as a functon the twist angle $\theta$.
Black dots are the total amplitude, and  the red squares and blue triangles are the intralayer and 
interlayer contributions, respectively.
Dotted curve represents the low-angle expression Eq.\ (\ref{eq_U_ee_low}) multiplied by the factor 0.6,
and the dashed horizontal line indicates the exact value in the high-angle limit, 3.8 ${\rm meV \cdot nm^2}$.
}
\label{fig_u_ee}
\end{figure}

\subsection{Phonon-mediated electron-electron interaction}

The phonon-mediated electron-electron interaction is written as
\begin{align}
& H_{\mbox{\scriptsize e-e}} =
\frac{1}{S} 
\sum_{\sigma\sigma'} \sum_{\alpha \alpha' \beta \beta'} \sum_{\Vec{q} \Vec{k} \Vec{k}'\in{\rm MBZ}} 
c^{\dagger}_{\beta'\Vec{k}'-\Vec{q}\sigma'} c^{\dagger}_{\beta\Vec{k}+\Vec{q}\sigma} 
c_{\alpha\Vec{k}\sigma} c_{\alpha'\Vec{k}'\sigma'} 
 \nonumber\\
& \hspace{10mm} 
\times U_{\mbox{\scriptsize e-e}}^{(\sigma,\sigma')}
 (\beta',\Vec{k}'-\Vec{q}; \beta,\Vec{k}+\Vec{q}; \alpha,\Vec{k}; \alpha',\Vec{k}'),
\end{align}
where $\sigma, \sigma'$ represent the spin-valley degree of freedom
which was omitted above, and we defined
\begin{align}
&U_{\mbox{\scriptsize e-e}}^{(\sigma,\sigma')}
 (\beta',\Vec{k}'-\Vec{q}; \beta,\Vec{k}+\Vec{q}; \alpha,\Vec{k}; \alpha',\Vec{k}')
 \nonumber\\
& 
 =
 \sum_n
 g^{(\sigma)}_{n,\Vec{q}} (\beta, \Vec{k}+\Vec{q}; \alpha, \Vec{k})
 g^{(\sigma')}_{n,-\Vec{q}} (\beta', \Vec{k}'-\Vec{q}; \alpha', \Vec{k}')\times
 \nonumber\\
& 
\frac{1}{2}
\left(
\frac{1}{
\varepsilon_{\alpha' \Vec{k}'\sigma'} - \varepsilon_{\beta' \Vec{k}'-\Vec{q} \sigma'} - \hbar \omega_{n,\Vec{q}}
}
+
\frac{1}{
\varepsilon_{\beta \Vec{k}+\Vec{q} \sigma} - \varepsilon_{\alpha \Vec{k} \sigma} - \hbar \omega_{n,\Vec{q}}
}
\right),
\end{align}
where $\varepsilon_{\alpha \Vec{k} \sigma}$ is the eigenenergy of state $|\alpha, \Vec{k}\rangle$
of the spin-valley sector $\sigma$. 
Similar to Eq.\ (\ref{eq_g_n_bar}),
we define the averaged interaction amplitude for the lowest two bands as
\begin{align}
\bar{U}_{\mbox{\scriptsize e-e}} 
\equiv
\frac{1}{N_\Vec{q}} \sum_{n, \Vec{q}\in{\rm MBZ}} \frac{1}{4} 
\sum_{\alpha,\alpha'=\pm}
\frac{|g_{n,\Vec{q}} (\alpha', \bar{\Vec{K}}+\Vec{q}; \alpha, \bar{\Vec{K}})|^2}
{-\hbar \omega_{n,\Vec{q}}},
\end{align}
which is obviously an attractive interaction. Here we neglected $\varepsilon_{\alpha\Vec{k}\sigma}$ in the denominator
to consider the small electronic band width in the low twist angles.

We can roughly estimate the magnitude of  $\bar{U}_{\mbox{\scriptsize e-e}}$ 
using the previous argument for the electron-phonon coupling $g_{n,\Vec{q}}$.
In the large angle regime, we replace $g_{n,\Vec{q}}$ with Eq.\ (\ref{eq_gq_low_angle}) 
and $\omega_{n,\Vec{q}}$ with $v_{\rm ph} q$, and obtain
\begin{align}
\bar{U}_{\mbox{\scriptsize e-e}}  \sim \frac{\gamma_0^2}{2\rho_r v_{\rm ph}^2} \sim 7.6 \, {\rm meV \cdot nm^2}.
\label{eq_U_ee_high}
\end{align}
In the low angle regime, Eq.\ (\ref{eq_g_intra_order}) and $\omega_{n,\Vec{q}} \sim v_{ph} /L_M$ lead to
\begin{align}
\bar{U}_{\mbox{\scriptsize e-e}}  
\sim \frac{\gamma_0^2}{2\rho_r v_{\rm ph}^2} 
(\Delta q \cdot L_M)^2.
\label{eq_U_ee_low}
\end{align}
Here the dimensionless factor $\Delta q \cdot L_M$ is given by
\begin{align}
\Delta q \cdot L_M \approx \frac{t_0}{\hbar v/a} \frac{1}{\theta}= \frac{2.7}{\theta {\rm [degree]}}. 
\end{align}
Therefore, the electron-electron interaction amplitude is enhanced in small twist angles. 

Figure \ref{fig_u_ee} plots the numerically calculated $\bar{U}_{\mbox{\scriptsize e-e}}$ as a functon the twist angle $\theta$,
where the black dots are the total amplitude, and  the red squares and blue triangles are the intralayer and 
interlayer contributions, respectively.
Here the dotted curve represents the low-angle expression Eq.\ (\ref{eq_U_ee_low}) multiplied by a factor 0.6,
and the dashed horizontal line indicates the exact high-angle limit, 3.8 ${\rm meV \cdot nm^2}$.
As expected, the intralayer component rises as nearly $\propto \theta^{-2}$ in the low angle regime.
The enhancement is suddenly interrupted around $0.8^\circ$,
and it is due to the band crossing between the lowest flat bands and the excited dispersive bands.
The interlayer contribution is also enhanced in the low twist angles
while not as much as the intralayer part,
and the total amplitude of the electron-electron interaction 
becomes as much as 80  ${\rm meV \cdot nm^2}$ at $\theta \sim 0.8^\circ$.
The characteristic energy scale of the phonon-mediated interaction is given by $\bar{U}_{\mbox{\scriptsize e-e}}/ S_M$
where $S_M=(\sqrt{3}/2)L_M^2$ is the moir\'{e} unit area.
At the magic angle $\theta \sim 1^\circ$, in particular,
$\bar{U}_{\mbox{\scriptsize e-e}}/ S_M$ is about 0.4 meV.
The dimensionless parameter for the interaction then becomes $\bar{U}_{\mbox{\scriptsize e-e}} D \sim O(1)$,
where $D$ is the density of states of the flat band 
which is typically a few $(S_M \cdot {\rm meV})^{-1}$. \cite{koshino2018maximally}
This indicates that the phonon-mediated interaction is strong in the nearly flat band.


\section{Conclusion}
\label{sec_conclusion}

We constructed a theoretical framework
to model the TBGs with lattice deformation and the electron-phonon coupling.
Starting from the tight-binding model, we write down the interlayer matrix element 
as a function of arbitrary lattice displacement  [Eq.\ (\ref{eq_umat_3d})],
and then obtain its long-wavelength continuum expression [Eq.\ (\ref{eq_umat_3d_approx2})].
The general formula Eq.\ (\ref{eq_umat_3d}) works for any twist angles
with arbitrary displacement vectors,  and a similar theoretical treatment would be applicable to
any two dimensional interfaces of van der Waals materials.
The long-range version, Eq.\ (\ref{eq_umat_3d_approx2}), has a simpler form and it is useful
to describe the low-angle TBGs with smooth lattice deformation.
We actually demonstrated that the lattice relaxation effect can be implemented into the Hamiltonian
by using Eq.\ (\ref{eq_umat_3d_approx2}), 
and the obtained model precisely reproduces the band structure of the original tight-binding model.
Finally, we applied Eq.\ (\ref{eq_umat_3d_approx2}) to the phonon problem,
and derived the matrix element between the electrons and moir\'{e} acoustic phonons.
Finally, we numerically estimated the electron-phonon coupling and phonon mediated electron-electron interaction
for the low-energy electronic states,
and found a significant enhancement in the low twist angles due to the superlattice hybridization.
While we focused on the long-range acoustic phonons, 
the electron-phonon coupling for the short wavelength vibrations (such as the optical phonons) 
can be described by starting from the general formula of Eq.\ (\ref{eq_umat_3d}).
We leave the detailed studies of these problems for future work.


\section{Acknowledgments}

M. K. thanks the fruitful discussions with  Debanjan Chowdhury. 
M. K. acknowledges the financial support of JSPS KAKENHI Grant Number JP17K05496. 

\bibliography{moire_cont_relax}

\end{document}